\documentclass[mathleft
]{an}
\usepackage{graphicx}
\usepackage{times}
\begin{document}

\Pagespan{789}{}
\Yearpublication{2006}%
\Yearsubmission{2005}%
\Month{11}%
\Volume{999}%
\Issue{88}%

\title{Searching for the missing baryons in the Warm-Hot Intergalactic Medium}

\author{X. Barcons\inst{1}\thanks{
  \email{barcons@ifca.unican.es}}}
\titlerunning{Missing Baryons in the WHIM}
\authorrunning{X. Barcons}
\institute{Instituto de F\'\i sica de Cantabria (CSIC-UC), 39005
Santander, Spain}

\received{24 Sep 2007} \accepted{11 Nov 2007} \publonline{later}


\abstract{%
  At low redshift ($z<2$), almost half of the baryons in the Universe are not found in bound structures like
  galaxies and clusters and therefore most likely reside in a Warm-Hot Intergalactic Medium (WHIM), as predicted
  by simulations.  Attempts to detect WHIM filaments at cosmological distances in absorption towards
  bright background sources have yielded controversial results that I review here.  I
  argue that a secure detection of absorption features by the WHIM is at the limit
  of the {\it XMM-Newton} capabilities, but feasible.  A proper characterisation
  of the whole WHIM belongs to the realm of future X-ray missions.}

\maketitle

\section{Introduction}

In the current cosmological paradigm, the content of the Universe is
dominated by the "dark sector" (Dark Matter and Dark Energy), with
only $\sim 4.5\%$ being in the form of baryonic matter.  This
distribution is observationally supported by the angular shape of
the Cosmic Microwave Background anisotropies and by counts of type
Ia Supernoave as a function of redshift. The small fraction of
baryons was inferred long ago by Big Bang nucleosynthesis being
responsible for the formation of light elements.  In particular a
higher baryon density early Universe would be unable to produce
enough Deuterium (and $^3$He) to the level required by its measured
abundance and a lower baryon density Universe would not generate
enough $^4$He. It is a remarkable success that the value of the
baryon density inferred from D/H measurements in the Lyman $\alpha$
forest towards distant QSOs agrees with that derived from the power
spectrum of the anisotropies in the CMB, with a value $\sim 4.5$\%
(for a Hubble constant $H_0=70\, {\rm km}\, {\rm s}^{-1}\, {\rm
Mpc}^{-1}$, a value used throughout this paper).

As noted in a number of works, at redshift $z>2$ the amount of
baryons found in the Universe is roughly consistent with the above
estimate (see, e.g., Fukugita, Hogan \& Peebles 1998). A significant
fraction of them reside in the Lyman $\alpha$ absorption systems,
and in particular in the Damped Lyman $\alpha$ Absorbers or DLAs. At
lower redshift, it is expected that the amount of baryons contained
in these absorbers decreases, as they should be forming stars.  The
number of absorbers per unit redshift decreases substantially
towards lower redshift in such a way that the amount of baryons
contained in the Lyman $\alpha$ absorbers is negligible at $z=0$.
However, the baryon density found in stars, interstellar and the
intra-cluster medium in the local Universe is significantly lower
than the 4.5\% of baryons contained in the Universe. This has been
summarized by Nicastro et al (2005a) with the conclusion that the
grand total of detected baryons at $z<2$ is $2.5\%$. Therefore this
opens a major cosmological issue:  45\% of the baryons in the local
Universe are missing.

On the other hand, it is theoretically inconceivable that all
baryons at low $z$ are locked into stars, galaxies and clusters of
galaxies. There are at least two reasons for this: first the
efficiency of forming galaxies in binding all baryons is surely not
100\%; second, the evolution of stellar populations in galaxies
involves Supernovae, winds and other phenomena that will end up
polluting and enriching the surrounding intergalactic medium. Using
simulations, it has been confirmed in a number of papers (Ostriker
\& Cen \break 1996, Cen \& Ostriker 1999, Dav\'e\ et al 1999) that a
significant fraction of baryons would reside today in a Warm-Hot
Intergalactic Medium (WHIM), with temperatures ranging between
$10^5$ and $10^7$K. Using cosmological simulations, Dav\'e\ et al.
(2001) concluded that this Warm-Hot component of the Intergalactic
Medium becomes more abundant towards low redshift, that it becomes
hotter with time and that its spatial distribution follows the
general Dark Matter filamentary structure. The WHIM is heated to
such high temperatures primarily by shocks, as assumed in the above
modeling. Extra heating from Supernovae or AGN of the intergalactic
baryons has also been invoked by Pen (1999) and Wu, Nulsen \& Fabian
(2001), in such a way that its thermal emission does not exceed the
extragalactic component of the soft X-ray background. The required
amount of heat deposited in the WHIM by these non-gravitational
processes (of the order of 1 keV per nucleon) will increase the
temperature of the intergalactic gas to $\sim 10^7$~K.

Most of the WHIM gas is contained in non-virialized structures.
Detailed simulations of the distribution of various ion species have
been performed by Fang, Bryan \& Canizares (2002), showing that higher ionisation species\\
tend to appear in higher density environments, but the abundant ions
such as OVII or OVIII are to be found in the low-density filaments,
and when observed in absorption towards distant X-ray sources might
give rise to an ``X-ray forest'' (Hellsten et al 1998). Velocity
broadening in these structures is significant ($\sim 100\, {\rm
km}\, {\rm s}^{-1}$).

In this paper I review some of the attempts conducted during the
last few years with the {\it XMM-Newton} and {\it Chandra} gratings
in attempting to detect WHIM filaments in absorption towards bright
background sources. A local WHIM component has been detected and
confirmed beyond any doubt.  However, the situation regarding
non-local WHIM filament detections is not fully settled. Earlier
claims of detections of absorption features at significant redshifts
were not confirmed later.  There is now a heated debate around the
detection of a couple of intervening WHIM filaments along the line
of sight towards Mrk 421, with unclear outcome.  After discussing
all this, I will suggest a possible way forward do detect WHIM
filaments with contemporary instrumentation.

\section{Detecting WHIM filaments}

According to simulations, the WHIM filaments potentially detectable
are constituted by extremely tenuous gas (density $\sim
10^{-4}-10^{-6}\, {\rm cm}^{-3}$) extending over vast regions of
space ($\sim 1\, {\rm Mpc}$). The temperature should exceed $10^5$K,
otherwise the usual UV absorption lines would be detected, and
cannot exceed too much $10^7$K as otherwise the CMB would be
significantly Comptonised.

The possibility of detecting the WHIM in emission has been
discussed, e.g., by den Herder et al (2004).  The intrinsic
difficulty in detecting tenuous gas resides in the fact that the
emissivity scales with the square of the density.  The figure of
merit for an extended source is the grasp (effective area times
solid angle times exposure time), and as discussed by den Herder et
al (2004) this requires a dedicated mission. High spectral
resolution across the field of view is essential to fight
efficiently the background, which is rich in emission lines. Indeed,
detecting the WHIM in emission would yield the most complete
picture, as in addition to redshift distribution, physical state of
the filaments, chemical abundances etc. it would also yield the
spatial distribution itself, i.e., the filamentary structure.

The alternative way of detecting tenuous gas is in absorption
towards a bright background source. For a given resonant transition,
the equivalent width of the absorption lines scales as the gas
density, as opposed to the gas emissivity which scales as the square
of the density. Therefore, absorption is in principle an easier way
to detect the low density WHIM. The most important requirement then
is that a bright enough background source must be found.  High
resolution spectroscopy (for a point source) is also essential to
detect weak absorption features, as the ones expected from tenuous
gas.  The obvious disadvantage with respect to the detection of
emission from the WHIM is that each observation traces only a single
line of sight and therefore information on the spatial distribution
is in principle lost (or it requires multiple observations along
nearby lines of sight).

There is, however, an important additional difference between the
gas seen in emission and absorption.  Emission dims as the square of
the distance and therefore samples preferentially the local WHIM
component.  The redshift dependence of the absorption equivalent
width is very mild (only through the stretching of the wavelength
scale, so the dependence with distance is marginal), so in principle
higher redshifts can be explored with absorption experiments.  The
difficulty resides, obviously, in finding bright enough targets at
significant cosmological distances.

At the spectral resolution of contemporary instruments like {\it
Chandra}/LETGS or {\it XMM-Newton}/RGS (with spectral resolution
slightly better than $1000\, {\rm km}\, {\rm s}^{-1}$), most WHIM
absorption lines, with a Doppler velocity parameter\\ $b\sim 100\,
{\rm km}\, {\rm s}^{-1}$ will be unresolved.  The minimal equivalent
width detectable in a properly sampled spectrum, with signal to
noise $S/N$ per spectral resolution element (of width
$\Delta\lambda$) is of the order of a few $ \Delta\lambda
(S/N)^{-1/2}$. Experience with UV and optical absorption line
spectroscopy towards distant QSOs demonstrates that unless $S/N$
exceeds several tens, the spectrum will be polluted with large
numbers of spikes which will make the detection of weak absorption
features almost impossible.  If that condition can be met, then the
minimum detectable equivalent width is a fraction of
$\Delta\lambda$.

In practice this means that both spectral resolution and effective
area are essential for this exercise. The most important transitions
used so far (OVI K$\alpha$ at 22.02\AA, OVII K$\alpha$ at 21.60\AA,
OVIII Ly$\alpha$ at 18.97\AA) occur at a wavelength around $20\,
$\AA\ which we take as a reference. The RGS on board {\it
XMM-Newton} has an effective area of $\sim 50\, {\rm cm}^2$ (in each
RGS unit), while the LETGS on {\it Chandra} has an effective area
$\sim 20\, {\rm cm}^2$ at this wavelength.  The spectral resolution
of both instruments are formally equal $\lambda/\Delta\lambda\sim
300$ in first order spectroscopy at 20\AA.  However, as discussed
thoroughly by Williams et al. (2006a), the RGS line spread function
has significant extended wings, which subsequently reduces the
sensitivity to weak absorption lines (see fig. 3 in that paper). As
a consequence we do not have at the moment an instrument that is
optimal for the detection of WHIM features in absorption: while {\it
XMM-Newton} has the capability of collecting the counts, {\it
Chandra} has a better suited spectral resolution.

Despite all this, there has been an important success in the search
for the WHIM.  Nicastro et al (2002), using {\it Chandra}/LETGS data
towards Pks 2155-304, along with data from {\it FUSE}, found clear
evidence for a local WHIM component. Specifically OVII K$\alpha$ and
NeIX K$\alpha$ were clearly detected, while OVIII K$\alpha$ and OVII
K$\beta$ were detected at lower significance.  This was accompanied
by the FUSE detection of at least two velocity components of OVI
K$\alpha$. Using {\it XMM-Newton}/RGS data, Rasmussen et al (2003)
confirmed that the detection of the $z=0$ OVII K$\alpha$ is
ubiquitous among extragalactic lines of sight, with clear detections
towards Pks 2155-304, 3C273 and Mrk 421.

The origin of this WHIM component is however still unclear.  Given
the resolution of the existing X-ray dispersive spectrometers, the
velocity and structure of the absorbing clouds cannot be determined
with enough accuracy to distinguish between an origin in the
interstellar medium, the Galactic halo or an extragalactic origin
associated to the Local Group.  This has been extensively discussed
by Nicastro et al (2002), Nicastro et al (2003) and Williams et al
(2006b) among others.  The mass implied in this absorbing gas
component is clearly dependent on the scale spanned, and in the most
extreme case of being associated to the Local Group it could account
for all the missing baryons in this structure; if, on the contrary,
the $z=0$ WHIM absorber is associated to the ISM or the Galactic
halo, the mass implied would be small.  As it has been extensively
demonstrated by many years of work on UV/optical QSO absorption
lines, and confirmed here, the importance of spectral resolution
when trying to assign absorbers to known objects (i.e., galaxies)
cannot be overstated.

\section{Detecting non-local WHIM}

Detecting WHIM absorption beyond the local group and trying to
somehow quantify the amount of baryons residing in this phase has
been and is still a major target of X-ray spectroscopic studies.
Ultimately we are dealing with the fate of almost one of every two
baryons present in the Universe, so it is hardly surprising that
various groups are devoting a lot of effort to this.

The strategy has been to target the brightest background sources
with the dispersive spectrometers on board {\it XMM-Newton} and {\it
Chandra}. For reference, an X-ray source with a 0.5-2 keV flux of
$10^{-11}\, {\rm erg}\, {\rm cm}^{-2}\, {\rm s}^{-1}$ requires
several hundreds of ks to produce a spectrum which has a $S/N$ of a
few tens per spectral resolution element, as required if weak
absorption lines need to be detected.  Of course, the sources need
to be at cosmological distances to sample a significant path along
the line of sight.  Often, the highly variable background sources
have been observed when in high state to maximise the $S/N$.

There have been various tentative detections of WHIM absorption
lines at significant redshift (Fang et al 2002,\\ Mathur et al 2003)
that turned out not being confirmed by deeper observations (Cagnoni
et al 2004, Rasmussen et al 2003).

The most recent and detailed work on this has been conducted on the
highly variable Blazar Mrk 421 ($z=0.03$).  Nicastro et al (2005a)
and Nicastro et al (2005b) claimed the detection of two WHIM
absorption systems towards this line of sight, using $\sim 200$ ks
of {\it Chandra}/LETGS data, when the source was at its historic
maximum. The WHIM filaments were reported at $z=0.011$ (with
detections of OVII and NVII at a combined significance of
3.5$\sigma$) and another one at $z=0.027$ (with detections of OVII,
NVII and NVI, at a combined significance of 4.9$\sigma$).  The
excitement that triggered these tentative discoveries was enhanced
by the estimates presented by Nicastro et al (2005b) whereby the
inferred amount of baryons in the WHIM, should these detections be
proven reliable, would be $\Omega_{\rm WHIM}\sim
2.7_{-1.9}^{+3.8}\%$ for an O/H abundance of 0.1 solar, i.e., the
WHIM would contain exactly the amount of missing baryons at low
redshift.

The same team tried to confirm later these findings with a stacked
{\it XMM-Newton}/RGS spectrum of the same source, collecting $\sim
500$ ks of good data. In the paper by Williams et al (2006a) it was
concluded that despite the long {\it XMM-Newton} exposure, the data
were not sensitive enough to either confirm or reject the {\it
Chandra} findings.  For this team, the main reasons for this lack of
sensitivity reside in the large number of narrow instrumental
features caused by bad detector columns, the intrinsic line spread
function (due to its extended wings) and the presence of fixed
pattern noise at 29\AA. Williams et al (2006a) found upper limits to
absorption features expected from the {\it Chandra} data around
4m\AA, while 2-3m\AA\ features were expected.

All this has been severely disputed by Rasmussen et al (2007).  This
competing team has analyzed a total of 950 ks of RGS data on Mrk
421, with a very careful and highly non-standard processing to
isolate many detector spatial and temporal features. In addition,
they claim that stacking data from a variable source should not be
done, but rather an analysis of the combined data set should be
performed.  The net result is that Rasmussen et al (2007) do detect
features associated with the local WHIM as weak as 2.5 m\AA, and
find upper limits to the features expected from the non-local WHIM
absorbers detected by {\it Chandra} at the level of 1.9m\AA\ (the
consistency between LETGS and RGS data has been addressed in detail
by Kaastra et al 2006). Rasmussen et al (2007) conclude that {\it
XMM-Newton}/RGS data do not confirm the existence of the two WHIM
filaments reported by Nicastro et al (2005a) and Nicastro (2005b).
Further, Rasmussen et al (2007) point out that the expected number
of absorbers towards Mrk 421 down to the best attainable sensitivity
is very small, in addition to the fact, remarked by all groups in
this field, that the O/H abundance in the WHIM is unknown and
difficult to predict from fundamental cosmological principles.

\section{A possible way forward}

The discussion presented in the previous section shows that the
detection of the WHIM features is at the limit of the capabilities
of contemporary X-ray dispersive spectrometers. Whether non-local
WHIM filaments have been detected is, to say the least,
controversial. Even if {\it Chandra} or {\it XMM-Newton} end up
detecting one or a few WHIM absorbers, the full characterisation of
the WHIM will have to await future missions, such as
XEUS\footnote{{\tt http://www.rssd.esa.int/XEUS}} which are capable
of collecting many more photons at a similar or better resolution
that the current X-ray spectrometers.

For the time being, my personal view is that the race to detect the
first WHIM absorbers and confirm them should be more
success-oriented and targeting the detection of a few of such
absorbers - to the best of the current knowledge.  I propose to keep
in mind three points:

\begin{itemize}

\item Conduct a ``serendipitous'' search, i.e., do not target specific
lines of sight with nearby groups or clusters or known
high-ionisation absorbers from the UV.  Only in this way, and
following an eventual detection, we might have a direct idea on how
many baryons are contained in the WHIM.

\item Go for a ``secure'' detection, i.e., sample enough line of sight
(with one or several targets) to span a long $\Delta z$ range, in
such a way that the number of expected absorbers is of several.

\item Metal abundance will always be a caveat, as it is very
difficult to reliably predict.  All the numbers will scale with the
value of O/H.

\end{itemize}

With all this in mind, I used the Chen et al (2003) predictions for
the number of OVII and OVIII absorbers per unit $z$ as a function of
minimum detectable equivalent width. I then adopted a conservative
3m\AA\ limit, which Rasmussen et al (2007) demonstrated that can be
achieved in an RGS spectrum that contains 10,000 counts in a 50
m\AA\ bin (this is formally S/N=100 per resolution element). In
terms of $dN/(d\ln (1+z))$, we can expect 4 and 3 OVII and OVIII
absorbers respectively, for a 0.1 solar metalicity WHIM.

Table 1 shows the RGS exposure time needed for a few distant
targets, at significant $z$ along with the number of expected
absorbers. Approximately, 10 Ms of {\it XMM-Newton} RGS exposure are
needed for a source with 1 ROSAT PSPC ct/s. Unfortunately only RGS1
counts can be used, as the region around the detectable Oxygen
features in the RGS2 is covered by one of the non-functioning CCDs.

\begin{table}
\centering
\caption{RGS exposure times and expected number of absorbers}
\label{tlab}
\begin{tabular}{lccc}
\hline
Target& $z$ & $t_{exp}$ (Ms) &$ N_{abs}$\\
\hline
3C 279 & 0.53 & 20 & 2.9\\
4C 71.07 & 2.17 & 14-19& $\geq 10$(*)\\
3C 454.3 & 0.86 & 16 & 4.3\\\hline
\end{tabular}

(*) The exact number depends on the precise ionisation history of
the WHIM.
\end{table}

The exposure times are very long, and still carry the uncertainty of
the metalicity. But exploring the fate of 50\% of the ordinary
matter in the Universe is among the most important challenges of
contemporary Astrophysics.

\acknowledgements I'm grateful to Frits Paerels, Yoh Takei and Andy
Rasmussen for illuminating discussions. Financial support for this
work has been provided by the Spanish Ministry of Education and
Science under project ESP2006-13608-C02-01.


\end{document}